# Remembrances of Michael E. Fisher

David R. Nelson
Lyman Laboratory
Harvard University
Cambridge Massachusetts 02138

Abstract

This contribution will be published in "50 years of the renormalization group", dedicated to the memory of Michael E. Fisher, edited by Amnon Aharony, Ora Entin-Wohlman, David Huse, and Leo Radzihovsky, World Scientific.  These personal remembrances come in three parts. The first contains a brief personal perspective on Michael E. Fisher's contributions to science. The second tells how I came to work with Michael and describes events while I was under his supervision during my graduate years at Cornell University. The third part summarizes recent work, done in collaboration with Suraj Shankar, on thermalized buckling of isotopically compressed thin (perhaps atomically thin) sheets of materials such as graphene or $MoS_2$. These investigations were inspired by Michael's beautiful work on the effect of constraints at critical points, with fluctuations at all length scales, which leads to "Fisher renormalization" of critical exponents. [1] Thin fluctuating sheets embedded in three dimensions, when they are tensionless as in a cantilever or "diving board" geometry, are automatically at a critical point everywhere in a low temperature flat phase. However, when we consider thin sheets supported on multiple sides in various ways, Fisher's ideas lead to the inequivalence of isotensional and isometric thermodynamic ensembles, which triggers dramatic differences in some of the critical exponents associated with the two types of boundary conditions. [2] Readers not interested in my experiences while a student at Cornell University may only want to read parts I and III. There are also some concluding remarks.

I.    Remarks on Michael Fisher's contributions to science

Most readers of this book will know that Michael Fisher, who when he passed away in November of 2021, was University Professor Emeritus and Regents Professor at the University of Maryland, and Horace White Professor Emeritus of Chemistry, Physics, and Mathematics at Cornell University, was a giant of statistical mechanics. Michael was in fact a towering influence for at least 60 years, in *many* fields, including condensed matter physics, statistical mechanics, applied mathematics, and physical chemistry. With Ben Widom, Leo Kadanoff, and Ken Wilson, he helped usher in the modern age of critical phenomena and condensed matter physics, with anomalous dimensions, scaling, and the epsilon expansion. He was a member of numerous honorary societies including the Brazilian Academy of Sciences, the Indian Academy of Sciences, and the Royal Norwegian Society of Sciences and Letters, to name just the more exotic organizations. His awards included the Irving Langmuir Prize in Chemical Physics, the Guthrie Medal, the Michelson-Morley Award, the Boltzmann Medal, and the Onsager Prize. In 1980, Michael, Leo Kadanoff, and Ken Wilson shared the Wolf Prize in Physics for their seminal contributions to understanding phase transitions and critical phenomena.

I would like to provide my personal perspective about Michael's striking contributions to science over the years. As many readers will know, during the final two decades of his scientific career, Michael did beautiful work on physical biology, which you will learn about elsewhere in this volume. Remarkably, physical biology *also* benefited considerably from his pioneering work on statistical mechanics in the 1960s. For example, Fisher demonstrated, via fugacity expansions, the existence of non-classical polymer critical exponents;[3] when applied to compute the bubble entropy for the classic Poland-Scheraga model of the helix coil transition,[4] one could finally understand the double-stranded DNA melting which underlies the polymerase chain reaction. Another of Michael's forays into statistical mechanics—the exact solution of the infinite spin one-dimensional Heisenberg model, [5] was later adapted by John Marko and Eric Siggia to predict the force extension curves in single molecule DNA stretching experiments, in the 1990s. [6]

In addition to these biophysics contributions, going back at least to the early1960s and moving onwards, Fisher's work on critical point correlations laid the foundations, along with Widom and Kadanoff, for the scaling theory of critical phenomena. The citation for his 1980 Wolf Prize (shared equally with Ken Wilson and Leo Kadanoff) reads in part

*Professor Michael E. Fisher has been an extraordinarily productive scientist, and one … at the height of his powers and creativity. Fisher´s major contributions have been in equilibrium statistical mechanics, and have spanned the full range of that subject. He was mainly responsible for bringing together, and teaching a common language to, chemists and physicists working on diverse problems of phase transitions.*

Fisher is uniquely responsible for the prediction of the anomalous critical exponent $\eta$ (Michael once owned a boat that sailed on Lake Cayuga in Ithaca, New York, named the Eta.), which controls the decay of order parameter correlations at the critical temperature. This exponent also plays a key role in quantum field theories, where it is closely related to the anomalous scaling dimensions. As Michael himself once said, the exponent $\eta$ is numerically small, but nevertheless quite important.

Perhaps Michael's most brilliant work was his invention, with Ken Wilson, of the epsilon expansion. [7] Nothing less than perturbation theory in the deviation of the dimensionality from four, and indirectly related to the concept of dimensional regularization and field theory, the epsilon expansion led, almost overnight, to a huge number of essentially exact solutions to previously intractable problems in static and dynamic critical phenomena. The relative stability of various fixed points explained the existence of universality classes; multi-critical points were introduced and classified; scaling functions were calculated; the dependence of critical exponents on dimensionality and symmetry was finally understood etc., etc. Although this revolution required the invention by Wilson of a new kind of renormalization group, Fisher's crucial idea of critical exponents in 3.99 dimensions is what brought this revolution to life.

It's interesting to recall how close Ken Wilson came to the $\varepsilon$-expansion before his pivotal collaboration with Michael. In his second paper laying down the foundations of his renormalization group ideas, [8] Wilson constructed an "approximate recursion formula" for the quartic Landau potential describing coarse-grained Ising spins after $\ell$ renormalization group transformations of the form

$$Q_\ell(y) = r_\ell y^2 + \lambda_\ell y^4. \tag{1}$$

With the help of an inspired wave packet analysis that thinned out the degrees of freedom with each renormalization group iteration by a factor 2 (Ken was very fond of powers of 2), he produced a renormalization group which led to recursion relations in $d$-dimensions for the coupling constants $r_\ell$ and $\lambda_\ell$ after $\ell$ iterations that read

$$\begin{aligned} r_{\ell+1} &= 4(r_\ell + 3q_\ell \lambda_\ell - 9q_\ell^3 \lambda_\ell^2) \\ \lambda_{\ell+1} &= 16 \times 2^{-d}(\lambda_\ell - 9q_\ell^2 \lambda_\ell^2) \end{aligned} \tag{2}$$

with $q_\ell = 1/(1+r_\ell)$. This transformation represents a thinning of the degrees of freedom corresponding to "Kadanoff blocks" of size $b = 2$. After an uncontrolled truncation, Ken then proceeded to solve his approximate recursion relation numerically on PDP-10 computer at Cornell's Wilson Synchrotron Laboratory and estimate critical exponents directly in three dimensions. However, in 20-20 hindsight, he could instead have noticed that the second entry in Eq. (2) can be rewritten, for a small quartic coupling constant $\lambda_\ell$, as

$$\lambda_{\ell+1} \approx b^{4-d} \lambda_\ell = b^\varepsilon \lambda_\ell, \quad \varepsilon = 4 - d \tag{3}$$

Thus, the nonlinear coupling constant shrinks with iteration for $d > 4$, but grows and approaches a nontrivial fixed point (now known as the "Wilson-Fisher fixed point") when $d < 4$! In 2017, I hosted Michael Fisher's David M. Lee Historical Lecture at Harvard, entitled "Ken Wilson, as I Knew Him", with Dave Lee in attendance, which he told me was likely the last public talk he would ever give. See Fig. 1.

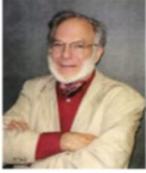

Fig. 1. Poster for Michael Fisher's 2017 historical lecture at Harvard University on his connections with Ken Wilson. In his lecture, Fisher described how he and Wilson arrived at the idea of critical phenonmena in 3.99 dimensions, inspired by how the physics of long self-avoiding polymers above four dimensions can be described as simple, noninteracting random walks. It was later pointed out by P. G. de Gennes that $\varepsilon = 4-d$ expansions for n-component spin systems can be adapted to understand the swelling of polymers with self-avoidance by taking the limit $n \to 0$.

In this lecture Fisher described in vivid terms how he and Ken were walking over to the Statler Faculty Club at Cornell University to have lunch after a seminar one day in 1971, when he pointed out that the physics of self-avoiding polymers simplifies to produce non-interacting random walk exponents above four dimensions and suggested the idea of a systematic perturbation theory in powers of $\varepsilon = 4-d$ for critical phenomena more generally. Ken Wilson produced the rough draft of their famous paper together a few days later. The final version of this paper[7] included not only the first epsilon expansion formula for the Ising model susceptibility critical exponent $\gamma$,

$$\gamma = 1 + \varepsilon/6 + O(\varepsilon^2) \text{ (Ising model)}, \qquad (4)$$

but also a beautiful analysis of two *coupled* Ising models, a kind of reformulation of Rodney Baxter's eight-vertex model.[9] Their analysis revealed that thermal fluctuations lead a *spontaneously generated* XY rotational symmetry at a critical point with the susceptibility exponent

$$\gamma = 1 + \varepsilon/5 + O(\varepsilon^2) \text{ (XY model)} \qquad (5)$$

Thus, the Wilson-Fisher $\varepsilon$-expansion illustrated not only the universality of critical exponents, but also revealed their dependence on the symmetry of the order parameter.

Michael's larger-than-life personality, a feature shared by all members of his family, led to pugnacious responses to various challenges. He was famous for standing up after talks at Joel Lebowitz's influential Statistical Mechanics Conference series (then at Yeshiva University; the 124$^{th}$ version took place at Rutgers University in May of 2023) and delivering blunt evaluations of talks he didn't like. (He could also deliver praise when he thought appropriate.) At a Cornell University talk by an experimentalist who claimed to have measured a violation of a critical exponent inequality near a critical point, Michael is alleged to have asked the hapless speaker if he really meant to overthrow the Second Law of Thermodynamics! Thorough cross-examinations of speakers at the stimulating weekly lunchtime seminars (blackboard only, no transparencies allowed) he organized with Ben Widom were an important feature of the education of graduate students and postdocs. Participating in these seminars encouraged (and probably expanded) my natural inclination to ask questions. Once, when I questioned a distinguished theorist from Canada at a neighboring seminar on superfluidity in Clark Hall, the speaker said he didn't answer questions from graduate students. There was a chorus of disapproval from the audience….Michael actively encouraged questions from everyone, including graduate students! Michael Fisher told me that he had a passion for raising the "level of hygiene" of the fields he entered. Those of us who learning the field of critical phenomena in the early 1970's were fortunate in at least two ways: (1) we had a new tool embodied in the renormalization group, a kind of calculus of functional integrals that few others knew, which enabled us to understand previously unsolved problems; and (2) we entered a field in which the issues and assumptions were clearly and precisely defined. Many of us discovered the importance of a high "level of hygiene" when we moved on to work in other areas of science, without the benefit of gadflies like Michael to clean up wrong ideas, sloppy thinking and misconceptions.

Michael Fisher's relentless challenging of authority often revealed itself when we met over the years. He insisted we ignore the warning signs, get out of a car driven by Eytan Domany and move quite close to a fierce-looking free-ranging buffalo, when we were together at a summer school in 1976, at the small Rocky Mountain town Banff, Canada. When I hosted Michael's 2017 Historical Lecture at Harvard, I took Michael to our faculty club for lunch. True to form, he ignored the velvet ropes deflecting us away from a fancy grand piano and took a straight line path to the relevant dining room. This action resembled his response to the velvet ropes blocking the stairs to the roof of the Church of Holy Sepulcher in Jerusalem, when he, my wife Pat and I visited during a Jerusalem Winter School I helped organized in 1988. This pushing past a barrier of velvet ropes took place during the 1st Palestinian uprising….Luckily, we weren't arrested; in fact, we got to meet the interesting Coptic Christians from Ethiopia who had been camping out on the roof for many years.

I now turn to a more personal account of my time during the early years of the renormalization group revolution at Cornell University, heavily influenced of course, by the presence of Michael.

II.   Interactions with Michael Fisher at Cornell University

It was my privilege to have Michael as my thesis advisor and postdoctoral mentor from the fall of 1972 to the summer of 1975. However, I almost didn't get to work with him.

How this happened is somewhat involved. I was in the fourth and final class of Cornell's Six Year Ph.D. program, an ill-fated experiment in elitist education, funded by the Ford Foundation, designed to take promising high school graduates, and help them get a Ph.D. at Cornell in only 6 years. I say "ill-fated", because many members of this program who entered Cornell, some as young as 15 or 16, were more concerned with learning how to drive or going on their first date than with dealing with the rigors of an accelerated academic program. As I recall, some had to leave Cornell before graduating and only 3/40 of the members of my class of "Phuds" got degrees from Cornell in the allotted time. Fortunately, I had already turned 18 (with some dating and driving experience…) when I arrived at Cornell in July of 1969 (we Phuds watched the first Apollo moon landing together in the basement of our dormitory), and this program turned out to be perfect for me. I was already in love with physics and mathematics, and the Six Year Ph.D. Program allowed me to construct a double-major in these subjects, with no complaints when I skipped various courses along the way. In addition to completing the requirements for a math major, I managed to take virtually every graduate physics course at Cornell in my first three years, with amazing teachers such as Ken Wilson, Hans Bethe, Bob Silsbee, N. David Mermin, Tung-Mow Yan and Peter Carruthers. When I took graduate statistical mechanics with Mark Nelkin, he told the class of an amazing Professor of *Chemistry*, Michael E. Fisher, who would teach a similar course with complementary perspective out of our Chemistry Department in a future year. Little did I know, I would eventually become a TA for this course!

As an undergraduate, I worked in the Clark Hall Science Library most evenings from 6pm to midnight. While wandering around this library in early 1972, I stumbled across the just-published Physical Review Letter by Ken Wilson and Michael mentioned above,[7] entitled *"Critical Phenomena in 3.99 Dimensions."* I was hooked and resolved to finish my final 3 years in the Six Year Ph.D. Program as Michael's graduate student. However, first I had to be formally admitted to Cornell's graduate Physics program. I had done well in virtually every graduate physics course Cornell offered and had overall grades that would allow me to become the salutatorian of my undergraduate class. My expenses as a graduate student, if accepted in the physics program, were guaranteed for 3 years by the Ford Foundation. I was thus somewhat disheartened to receive a polite letter from Cornell turning me down (*We receive many talented applicants and cannot accept everyone; we wish you all the best, etc.*). I showed my rejection letter to John Wilkins, my undergraduate physics advisor, who placed it on the desk of Bob Silsbee, the director of graduate admissions in Physics. I then left town on Friday for a three-day weekend, on a dubious trip to Kentucky (unrelated to my turndown) with my college roommate to purchase a bottle of "Rebel Yell" bourbon, a product only sold "South of the Mason-Dixon Line," due to unresolved issues associated with the U.S. Civil War. Fortunately for me, in addition to acquiring a bottle of Rebel Yell, things were straightened out about my admission by the following Monday: my rejection letter was intended for another David Nelson, who lived in Arizona. [10] I was formally admitted to graduate school in Physics, decided to stay at Cornell and was delighted when that fall Michael Fisher became my thesis advisor.

When I entered graduate school at Cornell in the fall of 1972, there were other exciting options. John Wilkins pointed out that the critical points of interest to Ken and Michael were merely a "set of measure zero" in the pressure-temperature phase diagram of most substances, and urged me to explore instead the beautiful experiments of Doug Osheroff, Bob Richardson and Dave

Lee, who had recently uncovered the remarkable superfluid phases of He$^3$ fermions in the basement of Clark Hall. I was in the lab with Doug Osheroff when some of their amazing data was coming in. I also vividly remember Bob Richardson presenting their latest results at an informal condensed matter physics lunch, where he revealed for the first time glitches in the time evolution of the pressure in their Pomeranchuk cell during compression and subsequent decompression. Bob eloquently replied to theorists insisting they had discovered the "Balian-Wertheimer phase" or the "Anderson-Brinkman-Morel" state, with something like "no, that feature is a reproducible pressure jump on our chart recorder whose explanation remains to be seen." This view of the primacy of experimental data over potential theoretical interpretations, however beautiful, made a deep impression on me. It was probably better for me that I *didn't* enter this exciting field. There are in fact *two* distinct superfluid phases of superfluid He$^3$, and it took quite some time to sort out all the subtle theoretical details, with significant contributions by brilliant experts like Tony Leggett [11]

And then there was of course Ken Wilson himself, who was already a legendary theorist in high energy physics. My graduate school roommate (who worked in high energy experiment) claimed that Ken's letter of recommendation for a Cornell faculty position from his Cal Tech professor Murray Gell-Mann read "Ken Wilson is the smartest person at Cal Tech except Feynman!" Feynman's letter supposedly said, "Ken is the smartest person at Cal Tech except me!" Ken had a self-effacing sense of humor – at an afternoon talk in the Laboratory of Nuclear Physics, there was a tremendous clap of thunder – he looked up at the sky outside and said calmly, "Did I say something wrong?" I wouldn't recommend trying this today, but Ken was rumored to have gotten tenure at Cornell without publishing a single paper while an assistant professor, on the very strong recommendation of Hans Bethe. However, I was already in love with statistical mechanics and, despite its close connections with quantum field theory (especially when formulated via path integrals), Michael Fisher seemed a better fit for me. In any case, I was extremely fortunate that my academic program allowed me to get to the frontiers of physics at Cornell already in 1972 – two future Nobel Laureates (Bob Richardson and Ken Wilson) were on my thesis committee. Dave Lee and Doug Osheroff, who were to share a Nobel Prize with Bob, were doing beautiful science in the basement of Clark Hall. Many of us felt that Michael Fisher and Leo Kadanoff (then at Brown University) should have shared the Nobel Prize with Ken[12]  --  I was thus surrounded by no less than five likely future Nobel laureates when I entered graduate school.

When I formally entered graduate school in 1972, Ken had recently returned from a stay at the Institute of Advanced Study at Princeton, where he gave a series of lectures on the Renormalization Group and the ε-expansion. These were published in Physics Reports with John Kogut. [13] Fortunately for me, Ken gave a course based on these lectures that fall, attended by students, postdocs and professors at Cornell. I was one of two graduate students who took Wilson's course for credit and was again the beneficiary (as I was when he taught me graduate level quantum mechanics) of his beautiful, inspirational lecture style. Since I was ready to do research, Michael suggested that I, in parallel, apply similar methods to the exactly soluble one-dimensional Ising model, where I could test fancy ideas about relevant and irrelevant variables, corrections to scaling and nonlinear scaling fields[14] in a simple context.  Michael was particularly intrigued about how the exact 1d renormalization group method could be adapted to handle antiferromagnetic interactions, a subject that would figure prominently in the rest of my thesis. In my investigations, I showed that integrating out every other spin mapped a 1d antiferromagnet onto a 1d ferromagnet after one iteration, but that integrating out 2 out of every three spins preserved antiferromagnetism. I presented this work that spring of 1973 at a seminar, with several luminaries present. I remain grateful to this day for the encouragement I received from David Mermin after that talk.

As I'm sure others will relate as well, Michael looked out for his students. In late May of 1973, Mel Green and others organized a wonderful conference to review the progress in understanding critical phenomena (and its implications for high energy physics) over the previous decade. The meeting was held in a Temple University Conference Center in rural Pennsylvania.[15] Michael used his persuasive powers (he was one of the organizers…) to arrange for me to attend, even though I was just a first year graduate student. [16] My roommate at that conference was Eric Siggia who came down from Harvard, where he was in the process getting a Ph.D. in only *five* years supervised by Paul Martin, without the benefit of a fancy program! This conference was an amazing opportunity for me, as I was able to hear, for the first time, a constellation of giants in the field expound on the ongoing revolution in statistical mechanics and quantum field theory, inspired by renormalization group ideas and the epsilon expansion. Experiments were represented by J.M.H. (Anneke) Levelt-Sengers from the National Bureau of Standards (now the National Institute of Standards and Technology) and also by Guenter Ahlers from Bell Labs. I remember Michael going out of his way to suggest that Guenter's talk on phase transitions in superfluid $He^4$ was likely to be one of the best talks by an experimenter I would ever hear. (It was….). He also strongly recommended the lucid, forthright talk of Michael Wortis on results from high temperature series expansions. I was introduced in addition to my future collaborator and postdoctoral mentor Bert Halperin by hearing him explain beautiful work on renormalization groups for dynamic critical phenomena carried out with Pierre Hohenberg and Shang-keng Ma, and remember Bert wondering whether we really understood the concept of a "metric" in the space of Hamiltonians when acted upon by Wilson's renormalization procedure. It was a pleasure to hear for the first time talks from those outside my Cornell acquaintances who would eventually become close personal friends and/or admired colleagues, such as Halperin, Shang Ma, Paul Martin, Leo Kadanoff, Franz Wegner, Eberhard Riedel, David Bergman, Daniel Amit, Jean Zinn-Justin, Masuo Suzuki, Edouard Brezin and Georgio Parisi. I still remember Edouard eloquently holding forth about the potential impact of renormalization groups on multiple fields of physics at one end of a long table at lunch, while Ken sat quietly eating at the other end.

Leo Kadanoff's talk had a particular impact on me, because he illustrated his seminal "Kadanoff block" renormalization group ideas by iteratively thinning the degrees of freedom of a 1d Ising model thus "decimating" out every other spin in a long chain and scooping me on my first year of graduate work. Michael spoke on my behalf to Leo, for whom I'm sure exact renormalization group transformations on 1d Ising models were a relatively minor development. Leo graciously encouraged Michael and me to go ahead and publish, although we did cite his Temple University talk in our eventual publication. [17] I was deeply impressed with Kadanoff, and ultimately had the privilege (after I left Cornell to become a postdoc) of collaborating with him, while he was still at Brown University, on an investigation with Jorge Jose' and Scott Kirkpatrick that ultimately became one of the most cited papers on our publication lists. [18]

My first year as a graduate student was not without other mishaps which required the assistance of Michael Fisher. When I took the oral exam that would allow me to officially start my thesis work, Bob Richardson, Ken Wilson and Michael were on the committee. I thought I could answer most questions they would throw at me and was reasonably confident. However, Bob Richardson, probably because he thought theorists needed a healthy dose of experimental reality, was out to get me. I *was* able to answer Ken's question about estimating the ground state energy of the hydrogen atom using the uncertainty principle. But then, after I explained Leo Kadanoff's ideas about the operator product expansion (independently invented by Ken in the context of quantum field theory), Bob asked me how a television worked. Well, I had wanted to become an electrical engineer during most of my high school years – although I ended up at Cornell, I applied to and was accepted by MIT for undergraduate studies for this reason. I had replaced the vacuum tubes on our home TV and I *knew* how a television worked….Undaunted by my correct answer, Bob tripped me up when he then asked me to compute the capacitance of a metal sphere…I panicked and failed miserably because I had no idea where to put the other plate. Ken Wilson kept saying "what's the only thing you can write down." The answer in MKS units is $C = 4\pi\varepsilon_0 R$, where $R$ is the sphere radius. Perhaps that this was only thing that *Ken* could have written down. However, there were *many* things I could have written down, especially because I didn't know that the other plate was at infinity! [19]

My committee decided that I would pass the exam, but with two important conditions, guided I'm sure by Michael: I had to serve as a teaching assistant for the section meetings of a freshman physics course in electricity and magnetism in the fall of 1973 (taught by Bobby Pohl and Boris Batterman), and I had to work for one semester in the laboratory of Watt Webb every Friday afternoon for a semester on an experiment on superfluid $He^3$-$He^4$ mixtures with an exceptional postdoc named Paul Leiderer. Teaching section lectures on electricity and magnetism forced me to really learn this subject, and taught me about teaching more generally. (Although most of the students liked my section lectures in their evaluations, one student asserted that "jumping up and down excitedly and waving my hands was a poor substitute for actually transmitting information."). he time in the laboratory studying dynamical critical phenomena near tricritcal points in $He^3$-$He^4$ mixtures with Paul Leiderer and Watt Webb led to my first Physical Review Letter[20], and was a truly transformative experience. Paul did most of the real experimental work with the background whine of a leak checker while hovering over a cryostat, and I helped mostly with aligning the apparatus and with the theory. Paul and Watt became lifelong friends; I connected with Watt's lab again when I returned to Cornell on sabbatical in 2004, which led to a biophysics theory paper on phase separation of lipid bilayers in giant unilamellar vesicles. [21]

My other indoctrination into experimental reality was a graduate course in experimental physics, required of theorists as well as experimenters at Cornell. (Harvard University has a similar requirement, instituted I believe by John van Vleck, who felt that theorists should not be on a "vitamin-free diet."). I still remember the thrill of finding a sharp singularity in the specific heat of a sample of beta-brass, a random alloy of copper and zinc. At a 50-50 concentration, beta brass orders like a two-sublattice antiferromagnet below an order-disorder transition. The head of that course, Paul Hartman, was very helpful and encouraging. He later wrote an engaging history of the Cornell Physics Department. I vividly remember learning two interesting things from Professor Hartman: (1) In the late 1930's, when he was a Cornell graduate student, he was worried about *his* oral exam. He went to Hans Bethe and said he knew most of the material but was worried because he hadn't really learned fluid mechanics. Bethe told him not to worry, because the physics faculty was getting rid of this subject, to make room for quantum mechanics! Sadly, fluid mechanics was long gone from U.S. physics curricula by the time I entered school -- I had to learn this subject on my own from a now battered copy of the text by L. D. Landau and E. M. Lifshitz. My copy still has squashed bugs from reading this text with the help of a kerosine lamp on a camping trip with my girlfriend, Patricia Schneider. (2) Paul Hartman also said the best Cornell student ever to take his legendary graduate course in graduate experimental physics was the famous theoretical physicist, Freeman Dyson. I was inspired by this remark to try some simple experiments later in my career. Reactions when I tried to get involved with experiments while on sabbatical at Brandeis with Bob Meyer and at Rockefeller University with Albert Libchaber ranged from lukewarm acquiescence to active discouragement! I finally succeeded in doing range expansion experiments on fluorescently labelled bacteria and Baker's yeast years later, while on leave from teaching at Harvard in the laboratory of Sharad Ramanathan (a former student of Michael's son, Daniel Fisher), in collaboration with Oskar Hallatschek, another theorist with a taste for experiments. [22] I shall always be grateful to Michael Fisher and Bob Richardson for insisting that I engage with experiment, and avoid becoming a "vitamin-free" theorist.

As mentioned above, some of the "Cornell crowd" were at that Temple Conference in the spring of 1973. Not surprisingly, given the recent breakthroughs in understanding superfluids and renormalization groups in Ithaca, this crowd included some exceptional postdocs! My friend, Amnon Aharony, was one of the Cornell postdocs at this meeting; I believe he reported on his beautiful work on phase transitions in spin systems with long range dipole-dipole interactions. Others postdocs from around the world (Amnon was from Israel) I knew at Cornell were Tim Padmore (USA), Joe Sak (Czechoslovakia), Roland Combescot (France), Monique Combescot (France), Michel Droz (Switzerland), Pierre Pfeuty (France), Bernie Nickel (Canada) and Alastair Bruce (Scotland). Some of them took me under their wing during the summer of 1972, when I shared a windowless office on the fifth floor of Clark Hall. They were generous in helping me deal with the Matsubara frequency sums in a Green's function calculation suggested by John Wilkins, before I signed on with Michael Fisher. But those were difficult times in condensed matter theory – the previous year there was only a single tenure-track job opening in this still-developing field in all of the United States, claimed by Joseph Straley (a former postdoc of Michael Fisher) at the University of Kentucky; at least one well-meaning postdoc asked me if I really wanted to go into this challenging field.

J. Michael Kostlerlitz showed up as a postdoc during my second year as a graduate student and talked about his beautiful paper[23] implementing ideas about topological phase transitions from his earlier work with David J. Thouless[24] and independent work by Berezinski. [25] We knew Mike was going to talk about his renormalization group solution of the famous two dimensional XY model, but it wasn't "our" renormalization group, i.e., the Wilsonian formulation involving local field theories, momentum shells and the epsilon expansion. Kosterlitz adapted a renormalization method developed for long range interactions in the Kondo problem by Anderson and Yuval [26] to understand virtually all important properties of XY models in two dimensions. Mike's talk was my first encounter with phase transition that had a *line* of fixed points, controlling an entire low temperature phase. Similar lines of fixed points (in fact, a whole surface of fixed points) played an important role in the two-dimensional melting transitions, with an intermediate fourth hexatic phase of matter, as I found when subsequently producing a theory of melting with Bert Halperin while a postdoc at Harvard. When Mike Kosterlitz and I were both at Cornell, we began to compete a bit to understand the then-fashionable subject of phase transitions with bicritical and tetracritical points. Michael Fisher stepped in and wisely insisted that we collaborate, which led to several papers together. [27] This important early connection with Mike helped us to collaborate effectively later (when I was a postdoc) on a paper that proposed a universal jump discontinuity in superfluid Helium films,[28] a prediction that played a role, via beautiful experiments by Dave Bishop and John Reppy at Cornell [29], in convincing people that 2-3 competing theories were wrong, and Kosterlitz and Thouless were correct with their ideas about two-dimensional superfluids and XY models of magnetism. We used renormalization group methods coupled with Josephson's scaling relation for superfluids in two dimensions to show that Mike's prediction of the critical exponent $\eta = 1/4$ describing the decay of correlations exactly at the phase transition implied a universal jump in the superfluid density, a prediction subsequently confirmed in a variety of experiments by Bishop, Reppy, Isadore Rudnick and others. [30]

My interactions with graduate students at Cornell were regrettably less extensive. Most of them were initially engaged with coursework, while I was trying to get started in research, although we did bond over of intramural games of softball and touch football on a team called the "Physics Zeroes." I remember breaking the nose of the captain of our football team in an unfortunate collision, and later playing shortstop in a softball game next to an open can of beer (as was also the case for other infielders and outfielders) in a so-called "beer game" against a competing physics team. However, I did make a close friend in H. R. Krishnamurthy, who arrived from India to enter in my year, to write an important thesis with John Wilkins and Ken Wilson about Ken's numerical approach to a renormalization group solution to the Kondo problem. [31]  One day, Krish walked into my office, concerned about a dispute between his two mentors, Wilkins and Wilson. John Wilkins always wanted more out of Ken, and said he wasn't satisfied with just the static susceptibility for the Kondo model, he wanted dynamics!  To which Ken Wilson replied with an obscene gesture. At least, I *think* it was an obscene gesture. When Kirsh tried to reproduce it, he used his index finger. I suggested that some other finger was likely involved. I also had enjoyable interactions with Eytan Domany, a tough Israeli graduate student who came to work with Michael shortly after me. Although Michael Fisher was most certainly Eytan's primary thesis advisor, I had the pleasure of working with Eytan somewhat in the capacity of a mentor during my final year.  Eytan and I used a renormalization group trajectory integral method I had helped develop to calculate equations of state for bicritical points. [32,33] As I remember our interactions, Domany initially thought most of the things I said were wrong….I was forced to work hard to explain my claims properly to him, a very valuable experience. Eytan became a close friend and was later the sponsor of numerous sightseeing adventures for me in Israel. This experience helped me when I later mentored graduate students while a postdoc at Harvard University, where I had the pleasure of working a bit with Bob Pelcovits and Daniel Fisher, when they were officially the students of their true thesis advisor, Bert Halperin.

The method Eytan and I used in our work together was an outgrowth of an accidental encounter I had over the telephone with Joseph Rudnick (the son of the UCLA experimenter Isadore Rudnick, mentioned above). While relatively isolated from the latest renormalization group developments as a postdoc at the Technion in Israel, Joe had done some remarkable diagrammatic renormalization group calculations. [34] He was now back in the US, and was phoning Michael Fisher, who was never in his office when Joe called. My tiny office (again, windowless) was then in the Baker Laboratory of Chemistry, next to Michael. Finally, with Joe again on the phone, Michael's frustrated assistant, Lynn Rabenstein, handed me the phone and said, "here,YOU talk to him!" Joe and I hit it off, discovered we had very similar interests and quickly wrote a paper together,[35] before ever meeting in person. I knew Joe was exceptional when I showed our draft paper to Ken Wilson, who said "Joe Rudnick, there's a smart guy." Ken had been a referee for one of Joe's papers. Joe eventually came to Cornell to give a talk, which led to our second paper together. [36]

In addition to our work on exactly soluble renormalization models in one dimension, my thesis with Michael Fisher concerned layered antiferromagnets called metamagnets[37], as well as phase transitions at bicritical and tetracritcal points. [38] We briefly thought that our metamagnet calculations could be adapted to demonstrate that Ising magnets and liquid-gas phase transitions were in different universality classes, until Ken Wilson pointed out to me that a crucial auxiliary field could be integrated out to show that the two theories were in fact equivalent. (The *dynamics* near these two types of critical phenomena is indeed different.) Michael and I were once talking about such problems in his office (which had a picture of Winston Churchill with an inspiring wartime quote), when he got an irate phone call from a colleague who claimed he had written an unfavorable letter of recommendation. Michael proceeded to pull out a copy of his (favorable) letter and read it out loud over the telephone! Thanks to Michael's support I managed to get my Ph.D. in 5.5 instead of 6 years. He employed me as a postdoc for the final semester of my last year, although I also had to help as a teaching assistant of his popular course on graduate statistical mechanics, offered by the Cornell Chemistry Department. Michael graciously gave me a lot of independence during my final year. Six of the papers originating from research I was involved in that year did *not* have Michael as a coauthor.

During my time at Cornell, Michael and Sorrel would have wonderful dinner gatherings at their home on the Parkway in Ithaca. (Ben Widom's house was just across the street). I remember meeting his sons Daniel and Matthew, who went on to become famous physicists themselves. I recall Matthew being very good at table tennis and persistently trying to get me to solve various mechanical puzzles. Daniel eventually came to me for advice on what physics courses he could skip when he entered Cornell at age 15. He would eventually become a valued colleague at Harvard, where he, Leo Radzihovsky and I played squash together, before Daniel was stolen away by Stanford. I brought Patricia Schneider (by then my fiancé) to the evenings at the Fisher home and was pleased to see that Michael clearly approved of her. Pat and I were married at Sage Chapel in Ithaca, New York (where we met singing in the choir), on a snowy day shortly after Christmas of 1975. Michael came to our wedding reception, where he explained the rules of English darts to my father, with a detailed follow-up letter. Our wedding present from Sorrel and Michael was a lovely chiming clock. This clock still sits today on the mantle above the fireplace in our living room. For 35 years or so after I became a faculty member myself, inspired by the generous hospitality of the Fisher family, Pat and I would invite all the stray postdocs and graduate students we could find, mostly international students from far away, to Thanksgiving dinners with our growing family at our home in Lexington, Massachusetts.

Thanks I'm sure to recommendations from Michael Fisher and Ken Wilson, I was selected as Harvard Junior Fellow, a three-year postdoctoral appointment that began in the summer 1975, a position that enabled me to work with Michael J. Stephen (then commuting to Harvard weekly from Rutgers University to be with his wife Johanna, on the faculty at the Harvard Medical School) to apply dynamical renormalization group methods to understand the large-distance and long-time properties of a randomly stirred fluid. Thus, I was able to capitalize on learning fluid mechanics during graduate student camping trips near Ithaca, New York, and eventually to work with two other important mentors besides Michael, first Leo Kadanoff and then Bert Halperin.

I now turn to some recent work inspired by Michael Fisher.

### III. Fisher Renormalization and thermalized buckling of isotopically compressed thin sheets

One of Michael Fisher's many contributions to critical phenomena prior to the invention of the $\varepsilon$-expansion was his 1968 paper [1] on the renormalization of critical exponents by hidden variables. What Fisher had in mind was, for example, an annealed density of mobile impurities in a background Ising model whose local density $c(\vec{r})$ can fluctuate, but whose overall number,

$$\int d^d r c(\vec{r}) = N_{imp}, \qquad (6)$$

is conserved, thus providing a constraint on the statistical mechanics. Fisher assumed the impurity density led to an additional contribution to the Hamiltonian $H$ of an Ising-like system with an "energy" coupling, say, to the square of the Ising order parameter. The constraint of course drops out if there are no impurities, $N_{imp} = 0$. In fact, such constraints will in general merely shift the location of the critical point a bit if the ensemble becomes grand canonical in the impurities, so that there is an impurity chemical potential $\mu_{imp}$, which plays the role of a Lagrange multiplier enforcing the constraint on average in the statistical mechanics, [39]

$$H \to H - \mu_{imp} \int d^d r c(\vec{r}), \qquad (7)$$

Here, $H$ already includes the coupling between the impurities and the Ising degrees of freedom. Fisher was able to show that an ensemble imposing a rigid constraint of a nonzero impurity concentration *changed* the path of approach to the critical point, so that the effective critical exponents close to $T_c$ such as $\alpha$, $\beta$ and $\gamma$ of the specific heat, magnetization and susceptibility respectively would change (in a slightly different notation from that used by Fisher) to [1]

$$\alpha \to \alpha' = -\alpha/(1-\alpha), \quad \beta \to \beta' = \beta/(1-\alpha), \quad \gamma \to \gamma' = \gamma/(1-\alpha). \qquad (8)$$

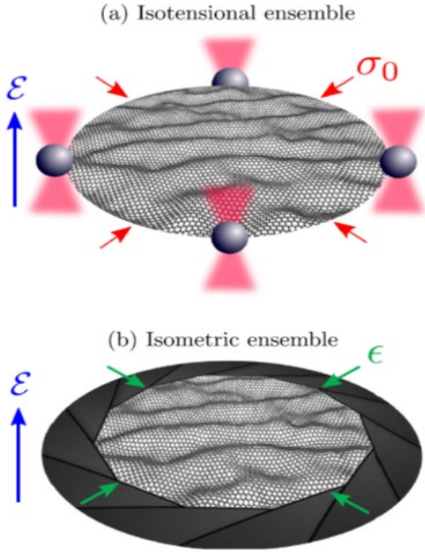

Fig. 2. A possible realization of two mechanical ensembles relevant, for example, to an atomically thin sheet of free-standing graphene, and not, as is sometimes the case, plastered to a 3d solid substrate that kills off flexural phonons by giving them a mass. (a) In the isotensional ensemble, multiple laser tweezers enforce a constant external stress applied to the membrane edge (in the spirit of single molecule biophysics experiments), while the edge boundary displacement fluctuates. (b) The isometric ensemble instead implements a clamped boundary condition used more often in experiments, with the external load imposed via a strain imposed on, say, a graphene membrane suspended across a hole of a fixed, but tunable size. Figure taken from Ref. [2], to which the reader is referred for more details.

Michael's paper, which has received over 1000 citations, is extremely important conceptually. Normally, we expect statistical mechanical averages to be *independent* of the precise statistical ensemble in the large number of degrees of freedeom, $N \to \infty$ thermodynamic limit of large system size. However, the standard proofs of ensemble equivalence[40] assume linear system dimensions $L \sim N^{1/d}$ much greater than the correlation length $\xi$, $L \gg \xi$. Michael realized that this assumption fails near a critical point, where there are fluctuations at all scales up to a diverging correlation length. Fisher's conclusions were subsequently confirmed via a renormalization group calculation in $d = 4 - \varepsilon$ dimensions by Rudnick et al.,[41] who found the exponent changes are reflected in a crossover from an unstable Wilson-Fisher Ising fixed point to a new stable fixed point with Ising-like Fisher renormalized critical exponents. (A related Ising-to-Ising fixed point crossover occurs in metamagnets.[37])

Although conceptually important, Fisher renormalization typically produces only small changes in actual exponents, since the specific heat exponent $\alpha$ appearing in Eq. (8) is almost always very small. [42] Recently, however, Suraj Shankar and I studied related Fisher renormalization effects for critical exponents near the finite temperature buckling transition of isotopically compressed thin sheets of materials such as graphene or MoS$_2$. [2] Here the inequivalence of isotensional (laser tweezers apply a fixed force at the edge) and isometric ensembles (here, a given hole size enforces a fixed strain) for thin supported membranes leads to dramatic differences between these two different boundary conditions for membranes, see Fig. 2.

Although experiments with atomically thin monolayers like graphene typically suspend the elastic sheets across fixed-size holes [43,44], it is useful to imagine a *variable* aperture size that could in principle be tuned by a camera shutter mechanism to change the strain isotropically, and possibly adjust it to zero. An external symmetry breaking electric field *E* perpendicular to average plane of the sheet could also be applied in either ensemble to bias the direction of buckling for charged membranes. In the absence of an external electric field, and in situations where boundary constraints allow easy relaxation to the tensionless state, such systems are *automatically* at a critical point at any point in the low temperature flat phase. [45] For example, when ribbons are in a cantilever geometry, [46] one end is free and the geometry resembles an extremely thin diving board where macroscopic strains can easily relax. Such systems have fluctuations at all length scales without the need for fine tuning, like long self-avoiding polymer chains in a good solvent, [47] and a collection of problems often called "self-organized criticality." [48]

To see how striking differences between isotensional and isometric ensembles can arise for thin, supported membranes, consider a long-wavelength free energy associated with a distorted elastic sheet characterized by in-plane phonon fields $u_1(x_1, x_2)$ and $u_2(x_1, x_2)$, as well as an out-of-plane "flexural" phonon field $f(x_1, x_2)$, see Fig. 3.

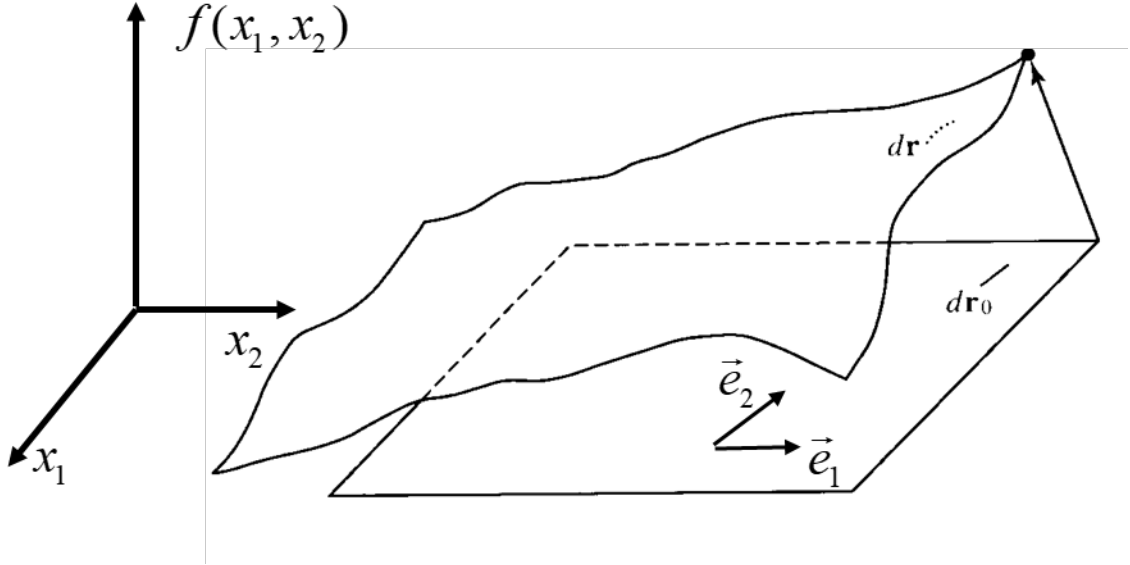

Fig. 3. Three-dimensional displacement vector from the upper right corner of a flat membrane reference state to a distorted state in three dimensions induced by, say, thermal fluctuations with both conventional in-plane phonon displacements and an out-of-plane flexural phonon field. See text for details.

Points in the original flat reference state, $\vec{r}_0 = x_1 \vec{e}_1 + x_2 \vec{e}_2$, are now mapped into three dimensions, via

$$\vec{r}_0 = x_1 \vec{e}_1 + x_2 \vec{e}_2 \rightarrow \vec{r}(x_1, x_2) = \vec{r}_0 + \begin{pmatrix} u_1(x_1, x_2) \\ u_2(x_1, x_2) \\ f(x_1, x_2) \end{pmatrix} \quad (9)$$

The effect of this mapping is to distort every line element $d\vec{r}_0 \rightarrow d\vec{r}$ with a new length given by

$$dr^2 = dr_0^2 + 2u_{ij} dx_i dx_j \equiv g_{ij} dx_i dx_j,$$
$$g_{ij} = \delta_{ij} + 2u_{ij} \quad (10)$$

where deviations from the identity $\delta_{ij}$ of the metric tensor $g_{ij}$ associated with the distortion is given by the nonlinear strain matrix

$$u_{ij}(\vec{x}) = \frac{1}{2}\left[\frac{\partial u_i(\vec{x})}{\partial x_j} + \frac{\partial u_j(\vec{x})}{\partial x_i} + \frac{\partial f(\vec{x})}{\partial x_i}\frac{\partial f(\vec{x})}{\partial x_j}\right] \tag{11}$$

(There are also higher order terms in the in-plane strains $\partial_i u_j$, but these turn out to be irrelevant compared to the terms kept here.) In the presence of edge forces and a symmetry-breaking field, the elastic energy of the distorted state is given by [49]

$$E = \frac{1}{2}\int d^2x \,[\kappa(\nabla^2 f(\vec{x}))^2 + 2\mu u_{ij}^2(\vec{x}) + \lambda u_{kk}^2(\vec{x}) - Ef(\vec{x})] - \oint_C d\ell\, \hat{v}_i \sigma_{ij}^{ext} u_j. \tag{12}$$

Here, the Lamé elastic parameters parametrizing the stretching energy are $\mu$ and $\lambda$, and $\kappa$ is a bending rigidity that penalizes deviations from the flat state. The final boundary integral is the work done by an external stress $\sigma_{ij}^{ext}$ with $\hat{v}_i$ being the outward unit normal (within the plane) to the boundary curve $C$. The final term in square brackets represents a potential energy proportional to an external out-of-plane electric field $E$ which couples directly to the height of a charged membrane. Note that the nonlinear term in Eq. (11) for the strain can be viewed as a matrix vector potential created by the flexural phonon field,

$$A_{ij}(\vec{x}) = \frac{\partial f(\vec{x})}{\partial x_i}\frac{\partial f(x)}{\partial x_j}. \tag{13}$$

Because the symmetric $2\times 2$ matrix $A_{ij}(\vec{x})$ has three independent components, whereas the in-phonon field field $\vec{u}(\vec{x})$ in Eq. (11) (at finite wavevectors) has only two components, flexural phonon distortions lead to geometrical frustration in the statistical mechanics whenever there is a non-zero Gaussian curvature.[50]

To see why the statistical membranes of thin elastic membranes with a shear modulus is like a system at its critical point (or a massless field theory), it is helpful to first set $E = 0$ and $\sigma_{ij}^{ext} = 0$ in Eq. (12) integrate out the in-plane phonon degrees of freedom $u_1(\vec{x})$ and $u_2(\vec{x})$ in the partition function. These quantities only occur quadratically in the in the membrane energy function. The resulting Gaussian functional integrals lead to and effective free energy $F_{eff}$, where[45]

$$\begin{aligned}F_{eff} &= -k_B T \ln\left(\int D\{u_1(\vec{x})\}\int D\{u_2(\vec{x})\} e^{-E/k_B T}\right)\\&= \frac{1}{2}\kappa \int d^2x\left[(\nabla^2 f)^2\right] + \frac{1}{4}Y\int d^2x\left[P_{ij}^T(\partial_i f \partial_j f)\right]^2 \equiv F_0 + F_1\end{aligned}. \tag{14}$$

Here, $P_{ij}^T = \delta_{ij} - \dfrac{\partial_i \partial_j}{\nabla^2}$ is the transverse projection operator and $Y = \dfrac{4\mu(\mu+\lambda)}{2\mu+\lambda}$ is the Young's modulus of the membrane. To better understand Eq. (14), note that if we represent the membrane position in space via the Monge form as $(x_1, x_2, f(\vec{x}))$, the unit normal is given by

$$\hat{n}(\vec{x}) = \dfrac{1}{\sqrt{1+|\vec{\nabla}f|^2}} \begin{pmatrix} -\partial_1 f(\vec{x}) \\ -\partial_2 f(\vec{x}) \\ 1 \end{pmatrix}. \tag{15}$$

The physics of a two-dimensional membrane embedded in three dimensions in its low-temperature flat phase in fact resembles somewhat a collection of interacting classical Heisenberg spins in two dimensions, with, however, genuine long range order in the normal field $\hat{n}(\vec{x})$, which plays the role of a 3-component spin vector. [45] See Refs. [51] and [52] for discussions of how and why elastic membranes are able to evade the famous Hohenberg-Mermin-Wagner theorem prohibiting a continuous broken symmetry and long range order in two-dimensions. The first term of the bottom line of the effective free energy in Eq. (14), $F_0 = \dfrac{1}{2}\kappa \int d^2x \left[ (\nabla^2 f)^2 \right]$, describes interactions that tend to align neighboring tipping vectors $\delta \vec{n}(\vec{x}) \sim -\vec{\nabla} f(\vec{x})$ of the normal away from the direction of long range order, which we take to be the $z$-axis. A Heisenberg-like exchange coupling $\kappa$ favors alignment. Thus, Eq. (14) is like a massless field theory in the tipping vector $\delta \vec{n}(\vec{x})$, with, however, an unusual nonlinear term $F_1 = \dfrac{1}{4}Y \int d^2x \left[ P_{ij}^T (\partial_i f \partial_j f) \right]^2$ quartic in the tipping vector. It can be shown that

$$-\nabla^2 \left( \dfrac{1}{2} P_{ij}^T (\partial_i f \partial_j f) \right) = \det\left( \dfrac{\partial^2 f}{\partial x_i \partial x_j} \right) = G(\vec{x}) \tag{16}$$

where $G(\vec{x})$ is the Gaussian curvature of the membrane at position $\vec{x}$. This observation leads to a long range interaction proportional to the Young's modulus $Y$ between distant membrane Gaussian curvatures, [50] which is one way of understanding how elastic membranes evade the Hohenberg-Mermin-Wagner prohibition against a continuous broken symmetries in two dimensions.

The profound effect of the nonlinear term in equation (14) is already evident in perturbation theory. A straightforward low-temperature spin-wave-like calculation in perturbation theory to lowest order in $Y$ gives for the renormalized wave-vector-dependent bending rigidity[45]

$$\kappa_R(q) = \kappa + k_B T Y \int \dfrac{d^2k}{(2\pi)^2} \dfrac{[\hat{q}_i P_{ij}^T(\vec{k})\hat{q}_j]^2}{\kappa |\vec{q}+\vec{k}|^4} + .... \tag{17}$$

For a membrane of linear size $L$ in the limit of small external wavevectors $q$, we have

$$\lim_{q \to 0} \kappa_R(q) \approx \kappa[1 + 3(vK)k_B T / (32\pi^3 \kappa) + ...] \quad . \tag{18}$$

Here the Foeppl-von Karman number is $vK = YL^2/\kappa$, a dimensionless measure of the strength of the nonlinear coupling. At first sight, the perturbative result Eq. (18) looks reassuring: For atomically or molecularly thin membranes, the dimensionless factor $k_B T/\kappa$ is typically quite small (of order 1/40) even at room temperature. [45] The numerical factor $3/32\pi^3$ is also rather small. Unfortunately for this perturbation theory, however, the factor $vK$ in the correction term *diverges* with system size! Upon substituting the elastic constants for a 8.5 inch $\times$ 11 inch piece of writing paper, one finds $vK \sim 10^7$, while for a relatively small $10\mu \times 10\mu$ square of graphene $vK \sim 10^{12}$. Thus, ordinary perturbation theory simply will not work for elastic membranes of any appreciable size. Such infrared divergences are tailor-make for graphical resummations of the most diverging terms in perturbation theory,[50] or alternatively for Ken Wilson's renormalization group. Such methods applied to the renormalized bending rigidity of elastic membranes lead to a scale-dependent elastic parameter characterized by a universal critical exponent $\eta$ [53,54]

$$\kappa_R(l) \approx \kappa(l/l_{th})^\eta, \quad \eta \approx 0.82. \tag{19}$$

where $\ell$ is a length scale such as the system size. The renormalized Young's modulus $Y_R$ also becomes scale dependent, and is characterized by $\eta_u$, another universal critical exponent,[53,54]

$$Y_R(l) \approx Y(l_{th}/l)^{\eta_u} \quad \eta_u \approx 0.36 \tag{20}$$

Note that the elastic "constants" $\kappa_R(\ell)$ and $Y_R(\ell)$ are not in fact constant but in fact depend on the length scale $\ell$ whenever $\ell \geq \ell_{th}$, where $\ell_{th}$ is a thermal length scale that can be calculated (in the the spirit of the Ginzburg criterion for when to expect non-classical exponents near a critical point) by asking at what length scale does the correction in Eq. (18) become comparable to the quantity it is correcting. The result is [55, 56]

$$\ell_{th} = \sqrt{\frac{16\pi^3 \kappa^2}{3k_B TY}} \tag{21}$$

The theory sketched above, although fairly well understood already by the late 1980's, received new life with the pioneering 2015 graphene cantilever experiments in the group of Paul McEuen at Cornell University, illustrated schematically in Fig. 4.[46]

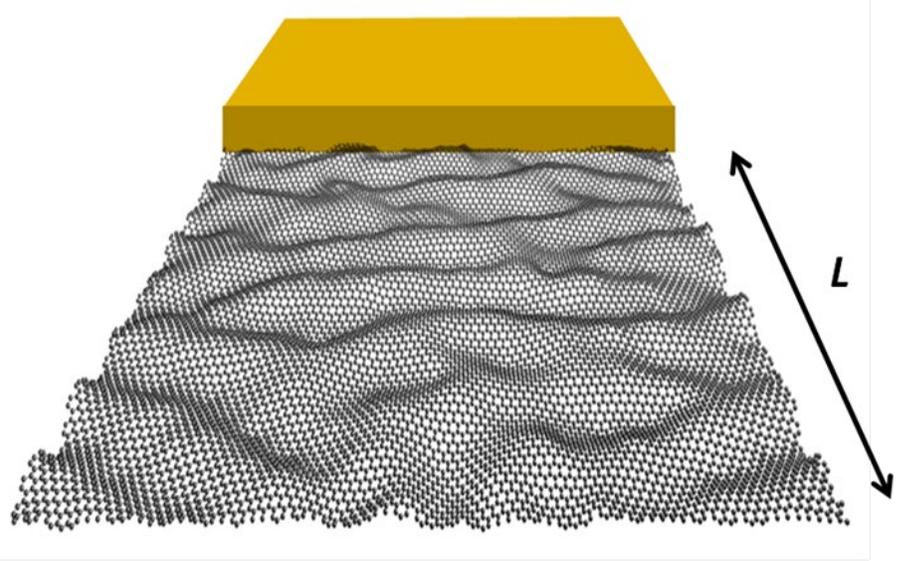

Fig. 4 Schematic of a thermally-fluctuating free-standing atomically thin graphene ribbon in a cantilever geometry, attached only at one end to a gold pad. After Ref. [46], with permission.

These authors pointed out that, using standard unrenormalized graphene elastic parameters, the thermal length at room temperature was only ~0.2nm. *All graphene experiments in materials above this tiny size will have strongly scale-dependent elastic parameters.* Although Blees et al. did not systematically vary the *width* of their free-standing graphene ribbons to demonstrate a scale dependence, they did change the length and were able to measure (both by measuring deflections of a metal pad and by measuring thermal fluctuations) a thermally renormalized bending rigidity at room temperature that was approximately *4000* times larger than that expected from measurements and quantum density functional theory near T = 0. We stress that relaxed graphene cantilevers in the flat phase like those shown in Fig. 4 are *automatically* at a critical point, with universal critical exponents and thermal fluctuations at all lengths up to the system size.

What happens when we apply isotensional or isometric constraints to this critical system, i.e., what is the effect of going from experimental setups like Fig. 4 to setups as in Fig. 2? Just as there are two ways to move away from a conventional critical point (temperature-like and magnetic field-like perturbations), there are two ways to destroy the criticality of free-standing elastic membranes. Upon returning to the full energy expression of Eq. (12), these turn out respectively to be an external electric field $E$ applied normal to a charged membrane and an external strain or stress ($\sigma_{ij}^{ext}$) applied to the edge of the membrane. For the isotensional ensemble, we assume an isotropic external edge force[2]

$$\sigma_{ij}^{ext} = \sigma_0 \delta_{ij} \qquad (22)$$

with no further constraints on the zero modes of the displacements or strain. The in-plane boundary displacements can adjust freely when $\sigma_0 \neq 0$, where with our conventions $\sigma_0 > 0$ corresponds to a tensile stress that smooths out thermal fluctuations and $\sigma_0 < 0$ represent a compressive stress that (when large enough) can cause a buckling transition, see below. For the isometric ensemble (likely to be easier to achieve experimentally), suspending a membrane across a circular hole of radius $R$ is assumed to lead to a fixed radial boundary displacement $\Delta_C$ along the contour $C$ [2]

$$\vec{u} = \Delta_C \vec{v} \equiv \varepsilon R / 2 \qquad (23)$$

where $\varepsilon$ (not to be confused with $4-d$) is the fixed strain associated with the isometric ensemble. The displacement at the boundary (measured relative to the relaxed state) in this fixed strain ensemble is given. For $\varepsilon > 0$, thermal fluctuations will be suppressed, while for $\varepsilon < 0$, they can be enhanced due to a buckling transition.

With two relevant perturbations to the tensionless state introduced, we can now define critical exponents. We only mention here three such exponents, two of which have large ensemble-dependent changes. It has been known since the mid-18th century and the time of Euler that thin plates like those shown in Fig. 2 will undergo a buckling transition for $\sigma_0$ or $\varepsilon$ sufficiently large and negative. Although Euler originally studied the buckling of rods, very similar phenomena arise when one compresses thin circular plates. [57] As shown in Fig. 5, for $E = 0$, there is a symmetry-breaking buckling phase transition beyond a critical value of the stress $\sigma_c$ (isotensional ensemble) or a critical value of the strain $\varepsilon_c$ (isometric ensemble), such that the maximum height $H_0$ of the thermalized structure rises continuously near the buckling transition, much like an order parameter,[58]

$$\begin{aligned} H_0 &\sim |\sigma_0 - \sigma_c|^\beta, \quad \text{(isotensional)} \\ H_0 &\sim |\varepsilon - \varepsilon_c|^{\beta'}, \quad \text{(isometric)} \end{aligned} \qquad (24)$$

The response of the $H_0$ to the external field is given by

$$\begin{aligned} \chi &= \left.\frac{\partial H_0}{\partial E}\right|_{E=0} \sim |\sigma_0 - \sigma_c|^{-\gamma}, \quad \text{(isotensional)} \\ \chi &= \left.\frac{\partial H_0}{\partial E}\right|_{E=0} \sim |\varepsilon - \varepsilon_c|^{-\gamma'}, \quad \text{(isometric)} \end{aligned} \qquad (25)$$

We expect these critical exponents $\gamma$ and $\gamma'$ to be the same on both sides of the transition.

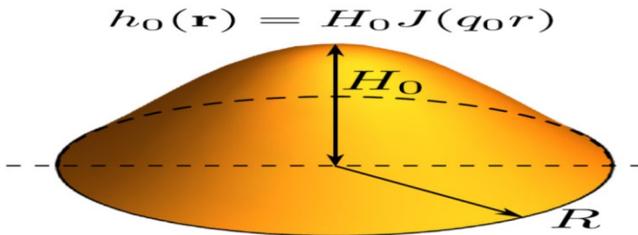

Fig. 5. Sketch of the first buckling mode with boundary conditions such that the height at the rim, for a circular plate of average radius R. There is an overall Bessel-function shape, and amplitude of the mode at the center of the circular frame, is corresponding to a small wave vector. Hinged boundary conditions are assumed at r = R in this example for simplicity. In the absence of a symmetry-breaking field, there is a broken symmetry, and states with height at the center are equally likely.

In fact, here unprimed exponents refer to the isotensional ensemble and primed exponents refer to the isometric ensemble. Our final critical exponents refer to the nonlinear response of $H_0$ to $E$ precisely at the buckling transition,

$$H_0 \sim E^{1/\delta}, \quad \text{(isotensional)}$$
$$H_0 \sim E^{1/\delta'}, \quad \text{(isometric)} \tag{26}$$

At zero temperature, these mean field Euler exponents are just [58], $\beta = \beta' = \frac{1}{2}$, $\gamma = \gamma' = 1$ and $\delta = \delta' = 3$. However, as shown using the Wilson-Fisher renormalization group (as well via the self-consistent screening approximation[54]) in Ref. [2], they can be dramatically different when one studies buckling transitions at finite temperature. The best current estimates are $\beta = \frac{1}{2}, \beta' = 0.718$ and $\gamma = 1, \gamma' = 1.436$, although there is no ensemble difference for $\delta$: $\delta = \delta' = 3$. (There is also no ensemble difference for the $\eta$-exponents associated with the running elastic constants: $\eta = \eta' = 0.821$, $\eta_u = \eta_u' = 0.358$.) Although the exponents $\beta$ and $\gamma$ are unchanged by thermal fluctuations in the isotensional ensemble, there are large differences in these quantities when one switches to the more experimentally accessible isometric ensemble. As mentioned above, the critical buckling thresholds in both ensembles are negative: $\sigma_c < 0$ and $\varepsilon_c < 0$. In fact, both these quantities vanish in the thermodynamics limit, $\sigma_c, \varepsilon \sim -1/R^2$, where $R$ is the membrane radius, as was known to Euler. [57] Thus, in the thermodynamic limit of infinite membrane size ($R \to \infty$), classical buckling is a thresholdless long-wavelength ($q_0 \sim 1/R \to 0$) instability, even though the buckling amplitude itself remains macroscopic in this limit. See Ref. [2] for a detailed exposition of the theory sketched above.

IV. Concluding remarks.

In this tribute to Michael E. Fisher, I attempted three things:

First, I touched upon his massive contributions to science and highlighted his larger-than-life, somewhat pugnacious personality. Michael's contrarian instincts were in full display in a provocative 1985 talk he gave at a symposium in honor of Niels Bohr at the American Academy of Arts and Sciences organized by Herman Feshbach, entitled "Condensed matter physics: does quantum mechanics matter?" Although I did not attend this meeting, Fisher explained to me in person the thesis in his talk that quantum mechanics was basically a side issue for most problems in the forefront of finite-temperature condensed matter physics at the time. He maintained that if the giants of 19<sup>th</sup> century physics, such as Boltzmann or Gibbs or Rayleigh, could be brought back to life in 1985, they could all be doing front line research in about a week, without taking time off to learn quantum mechanics. After all, to paraphrase Michael, in a Feynman path integral picture, say for massive bosons, most aspects of quantum mechanics can be captured just adding another dimension, of size $\beta\hbar$ ($\beta = 1/k_B T$), in an imaginary time direction arising from the quantum partition function.

With a pointer from Amnon Aharony, I was able to review Michael's contribution to the proceedings of this symposium. [59] Feshbach's original charge to Fisher asked for an "emphasis on any major unsolved problems (in condensed matter physics) and comments on any overlap with Bohr's ideas regarding the fundamentals of quantum mechanics." Feshbach's request clearly aroused Michael Fisher, much like waving a red cape at a bull!  Although Michael provided a wonderful review of the hot topics in condensed matter physics in the mid-1980's, such as the physics of polymer solutions, critical points in low dimensions, materials with quenched random disorder, quasicrystals, etc., he argues, whenever possible, for the irrelevance of quantum mechanical details, at least in the limit where finite temperature correlation lengths become large compared to a suitably defined length in the imaginary time dimension. Fisher of course accepted the importance of quantum mechanics for the collective behavior of most fermion problems, and he acknowledged that it underpins at a more microscopic scale most of the problems that interest him. I doubt Michael's ideas were well received at the Niels Bohr Symposium:  His 50-page paper, in which he must have invested considerable effort, has only been cited only 9 times in 35 years….

Of course, all physicists (including Michael) love the beauty of quantum mechanics. Two of my favorite courses at Cornell involved graduate quantum mechanics presented (in quite different ways) by Ken Wilson and Hans Bethe.  I myself have taught the standard two-semester sequence of graduate non-relativistic quantum mechanics frequently over the past 45 years. Nevertheless, I agree with Fisher that there can be advantages to stripping away some of the "quantum mysticism" by recasting quantum statistical mechanics as Feynman path integrals in imaginary time[60], full of classical world lines and commuting c-numbers. In the same vein, systematic coherent state path integral methods[61] allow us to describe $He^4$ superfluids at low temperatures in terms of a single macroscopic complex-valued wave function $\psi(\vec{r})$ that behaves classically upon coarse-graining. A middle road might be to slide back and forth between the two perspectives as needed. To me, with a background in soft matter physics as well as quantum condensed matter, the thermally excited (and possibly melted) Abrikosov flux lattices of wiggling quantized flux lines that appear in high temperature superconductors in a magnetic field[62] look like directed polymer melts! It can nevertheless be very helpful to pass via a transfer matrix technique to the equivalent quantum problem, which looks like a two-dimensional quantum superfluid at low temperature, albeit with a somewhat unusual ideal Bose gas boundary condition in imaginary time. [63] For a system of interacting, thermally excited vortex lines in 2+1 dimensions piercing a set of parallel $CuO_2$ planes in a high temperature superconductor, it can be shown that the analog of the long-sought-after quantum supersolid phase in two dimensions is all but *inevitable* at high magnetic fields. [64] Thick superconducting slabs, with microscopic cuprate layers ticking off discrete imaginary time intervals, map onto low temperatures in the equivalent 2d quantum system, with the physical temperature of the cuprate playing the role of Planck's constant. In a similar vein, it can be shown in this 2+1 dimensional world that the analog of a quantum *hexatic* phase is also quite likely, [65] an intermediate phase of matter with long range bond orientational order that seems to have been observed with diffraction off directed assemblies, not of actual Abrikosov flux lines in a superconductor, but of aligned, charged DNA polymers in 2 + 1 dimensions at room temperature, which have quite similar interactions. [66] Which is more interesting, *genuine* 2d quantum supersolids or quantum hexatics near T = 0, or analog systems of long directed polymer-like lines at finite temperatures in 2+1 dimensions? The answer seems to me to be a matter of taste. Another realm where the quantum ↔ classical analogy typically goes the other way is quantum antiferromagnets, where a hard quantum problem in d dimensions is mapped onto a better understood classical Heisenberg spin problem in a d+1 dimensional slab of thickness $\beta\hbar$. [67] Although this connection helped launch the exciting field of quantum phase transitions, [68] it *can* overlook subtleties like a Berry phase terms in the quantum action. [69] Overall, however, I think Michael Fisher would have *approved* of sliding back and forth between the quantum problems in d dimensions and classical analogs in d+1 dimensions, as needed to solve important problems. After all, Michael's favorite classical system, the two-dimensional Ising model, was solved by Onsager via a brilliant mapping of its transfer matrix onto the Hamiltonian of one-dimensional system of quantum fermions. [70] It is worth noting that that the Hamiltonians associated with transfer matrices for classical statistical problems can easily be *non-Hermitian*, as happens for the asymmetric six-vertex model. [71] Hence, the palette of Hamiltonians arising naturally from transfer matrices generated by classical statistical mechanics problems in one higher dimension (such as ice-models in a tilted electric field) is richer than the Hermitian problems associated with conventional quantum mechanics. Similar ideas led to a study of one-dimensional localization transitions in "non-Hermitian quantum mechanics," generated by tight binding models with asymmetric hopping matrix elements and diagonal disorder. [72] More recently, an imaginary time analog of the quantum Thouless pump has been studied for 2d directed polymers with repulsive interactions and interacting with a set of grooves in a 2d substrate. [73]

In the second part of this chapter, I tried to give a flavor of what it was like to work with Michael and be a student at Cornell in the early 1970's, an exceptionally exciting place and time. [74] I learned an amazing amount from the people around me, not only about science but also from exhortations such as "Drop everything when you have an opportunity to work with experimentalists!" or "Physical Review Letters is mere advertising, document your work in long papers in more humble journals!" via Michael. I also made lifelong friends such as Amnon Aharony, Eytan Domany, H. R. Krishnamurthy, Paul Leiderer and J. Michael Kosterlitz.[75]

In the third section, I gave a brief summary of recent work on the different critical exponents exihbited by isotensional and isometric ensembles for atomically thin elastic membranes.  In this work, Michael's prescient ideas about "Fisher renormalized critical exponents" from 1968[1] live on in modern context over a half century later. [2] (For a treatment of a related problem involving thermal buckling, see Ref. [76]). I hope Michael Fisher will also live on in Fig. 6.

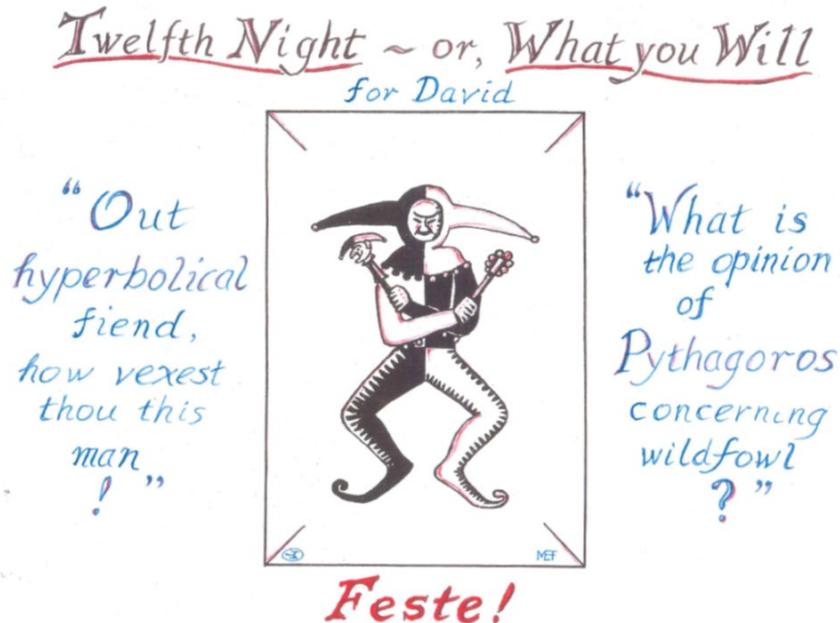

Fig. 6. Image created to advertise a college production of the Shakespeare play Twelfth Night by Michael E. Fisher. Gift of Michael for the 60th birthday of the author. Michael may have chosen to send this because the quote "Out hyperbolical fiend, how vexest thou this man!" conjures up work on geometrical frustration and the glass transition in spaces of constant negative Gaussian curvature.

The original came from an illustration by Michael for a college performance of the Shakespearean play Twelfth Night. To me, the image of the jester evokes Michael's catalytic role in science, whether it be defining and understanding critical exponents, inventing $\varepsilon$-expansions, questioning orthodox perspectives about the importance of quantum theory or just generally raising the "level of hygiene" for us all.

Acknowledgements

It is pleasure to acknowledge the fruitful collaboration with Suraj Shankar that led to Ref. [2] and Section III. I am also grateful for input about drafts of this chapter from Amnon Aharony, Eytan Domany, H. R. Krishnamurthy and Leo Radzihovsky. Any errors and possible omissions are, of course, my responsibility.

Prize Committee. Ken was surprised that Michael and Leo Kadanoff were not included in the award, as had been the case when they shared the Wolf Prize in Physics. After the Committee insisted that he receive the award on his own or not at all, he accepted and eventually called the Fisher home on the Parkway in Ithaca to let Michael know what had happened. Sorrel took the early morning call, but told Ken in no uncertain terms that she wasn't going to wake her husband from a sound sleep to tell him that he had *not* won the Nobel Prize! (Michael and Ken did talk later that morning, and all was well.) Leo Kadanoff's reaction on that day was to come to work at the University of Chicago in a full tuxedo with tails. When asked why, he replied "it's not every day that you *don't* win the Nobel Prize!"

[13] Wilson, K. G. and J. Kogut. "The renormalization group and the $\epsilon$ expansion." Physics Reports 12, (1974): 75-199.

[14] Wegner, Franz J. "Corrections to scaling laws." Physical Review B 5 (1972): 4529.

[15] Gunton, J. D., and M. S. Green, 1974, Eds., Renormalization Group in Critical Phenomena and Quantum Field Theory: Proceedings of a Conference, held at Chestnut Hill, Pennsylvania, 29–31 May 1973 (Temple University, Philadelphia).

[16] Michael drove some of us down from Ithaca, New York, to the Temple University site outside of Philadelphia. One of the *very* few times I ever saw him back down from authority is when he was stopped for speeding by a member of the Pennsylvania State Highway Patrol!

[17] Nelson, David R., and Michael E. Fisher. "Soluble renormalization groups and scaling fields for low-dimensional Ising systems." Annals of Physics 91, (1975): 226-274.

[18] José, Jorge V., Leo P. Kadanoff, Scott Kirkpatrick, and David R. Nelson. "Renormalization, vortices, and symmetry-breaking perturbations in the two-dimensional planar model." Physical Review B 16, (1977): 1217. This work was an investigation of XY models of magnetism with various symmetry-breaking fields, using a combination of duality transformations, renormalization group transformations and numerical work. My collaboration with Leo began when he was still at Brown University, in Providence, Rhode Island and I was a postdoc in the spring of 1976, in Cambridge, Massachusetts. I would drive down to Providence to see Leo, and he would drive up to Cambridge to see me. One day when working with Leo down at Brown, I suggested that we look up an important paper by J. Michael Kosterlitz, see Ref. [23] below. In those days, you had to search for papers in physical bound back issues of journals in the library. I was astonished to learn that Leo, after a decade on the faculty at Brown, had no idea where Brown's science library was located! We eventually found both the library and Mike's seminal paper, which built on earlier work with David J. Thouless. Only someone with Kadanoff's extraordinary creativity and ability to derive things from first principles could have functioned for 10 years at a major university without ever visiting the library!